\documentstyle[aps,12pt,epsfig,citesort,floats]{revtex}

\title{Decaying and kicked turbulence in a shell model}
\author{Jan-Otto Hooghoudt,  Detlef Lohse$^*$, and  Federico
 Toschi}

\address{
Department of Applied Physics  and J.\ M.\ Burgers Centre for
Fluid Dynamics, 
University of Twente, P.\ O.\ Box 217, 7500 AE Enschede, 
The Netherlands}

\begin{document}
\bibliographystyle{prsty}

\newcommand{\la}{\left\langle}
\newcommand{\ra}{\right\rangle}

\maketitle

\begin{abstract}
Decaying and periodically kicked turbulence are analyzed
within the GOY shell model, to allow for sufficiently large scaling
regimes. Energy is transfered towards the small scales in intermittent
{\it bursts}. Nevertheless, mean field arguments are sufficient to 
account for the ensemble averaged energy decay $E(t) \sim t^{-2}$ 
or the parameter dependences for the ensemble averaged total energy
in the kicked case. Within numerical precision, the inertial subrange
intermittency remains the same, whether the system is forced or decaying.

\end{abstract}

\section{Introduction}
Three dimensional
turbulent flows are characterized by a highly chaotic and intermittent
 transfer of energy  from the outer length scale
 down to the dissipative, inner length scale.
In this paper
we will focus on two different kinds of turbulent flows:
Decaying turbulence and periodically kicked turbulence.
Both types of flow have already been analysed within a mean field theory
\cite{loh94a,loh00}, but here we would like to focus also on intermittency
effects which cannot be described in the  mean field approach 
\cite{eff87}.

To be more specific:
With {\it decaying turbulence}  we mean a homogeneous and isotropic
 turbulent flow for which forcing is ceased
 from some time $t_0$ on.  
Therefore, eventually 
all the energy will be damped because of dissipative effects, but 
the  statistical properties of the 
decaying turbulent field are a priori not clear.
With periodically {\it kicked turbulence} we mean a turbulent 
field forced with short and very strong periodic pulses. 

The motivation of the paper is to study the effect of non-trivial 
forcing on the properties of the turbulence. 
Most numerical or theoretical studies assume a Gaussian random noise
forcing, acting on the largest scales only. 
But for most practical flows the forcing protocol is  obviously 
more complex 
and often periodic, be it pulsed flow through a pipeline or 
the earth's atmosphere 
driven by the periodical heating through the sun.
Another example of periodically kicked turbulence is 
the numerical realization of homogeneous shear flow 
\cite{pum96} where periodical remeshing is necessary. 
In between the kicks the turbulence is supposed to be freely 
decaying.
Therefore an understanding of decaying turbulence is 
required. Beyond this, decaying turbulence is of course
one of the classical examples of turbulent flow and an extended
literature exists, see e.g.\ 
\cite{hin75,my75,smi93a,eyi00,sta99}.

The aim of the paper   is to explore 
the statistical properties 
of turbulence both in the decaying and kicked case.
Therefore, we must have 
 excellent statistics. This turns out to be prohibative 
in a direct numerical simulation (DNS) of the Navier-Stokes equation and 
that is why we revert to study the problem in the context of 
the  GOY shell  model of turbulence \cite{bohr} in which a scaling 
regime of many decades can be achieved. To our knowledge the present
study is the first on the classical problem of decaying turbulence
with the help of a shell model.

Shell models are defined by a set of hierachically coupled ODEs for
the velocity modes which 
try to
 reproduce the physics of the energy flux of Navier-Stokes equation.
The dynamical equations for the GOY models read \cite{bohr,jen91,pis93,kad95}
\begin{equation}
\label{goyeq} \left({{d} \over {dt}}+\nu k_n^2\right)u_n = 
i k_n \left(a_n u_{n+2}u_{n+1} + b_n u_{n+1}u_{n-1} + 
c_n u_{n-1}u_{n-2} \right)^* + g \delta_{n,0}
\end{equation}
were $n=0,\dots,N-1$, $k_n=2^n k_0$,
$a_n=1$, $b_n= - \delta / 2$ and $c_n= - (1 -\delta)/4$ and the 
boundary conditions are $a_{N-1}=a_N=b_N=b_1=c_1=c_2=0$ in order for the GOY model to conserve the energy in the unforced and inviscid case ($g=\nu=0$). 
Traditionally, the free parameter $\delta$ is chosen to be $\delta=1/2$.
The values of the other parameter used were $N=22$, 
$g=(1+i)\cdot 10^{-2}$, and $\nu=10^{-6}$.

\section{Decaying turbulence: global properties}
The stationary (i.e. forced) simulations 
of eq.\ (\ref{goyeq}) 
were performed using fourth order Runge-Kutta
 with viscosity explicitely integrated. For the decay run we used the same algorithm but increased the time step keeping it $1/10$ of the dissipative time scale. During the decay process indeed the dynamics becomes slower and slower and investigation of long time properties is only possible using a scheme with an  adaptive time step.

We performed a very long stationary simulation of equation (\ref{goyeq})
 from which we collected an ensemble of starting configurations ($\sim 2500$ independent runs) for studying the free decay of GOY turbulence. 
The  starting configurations of the ensemble 
 were collected after some eddy turnover time in order to be statistically independent. We let each of the
 starting configurations 
decay according to  equation (\ref{goyeq}) (i.e. with $g=0$ and all other parameters unchanged).

\begin{figure}
\epsfxsize=.8\hsize
{\hskip 0.0cm{\centerline{\epsfbox{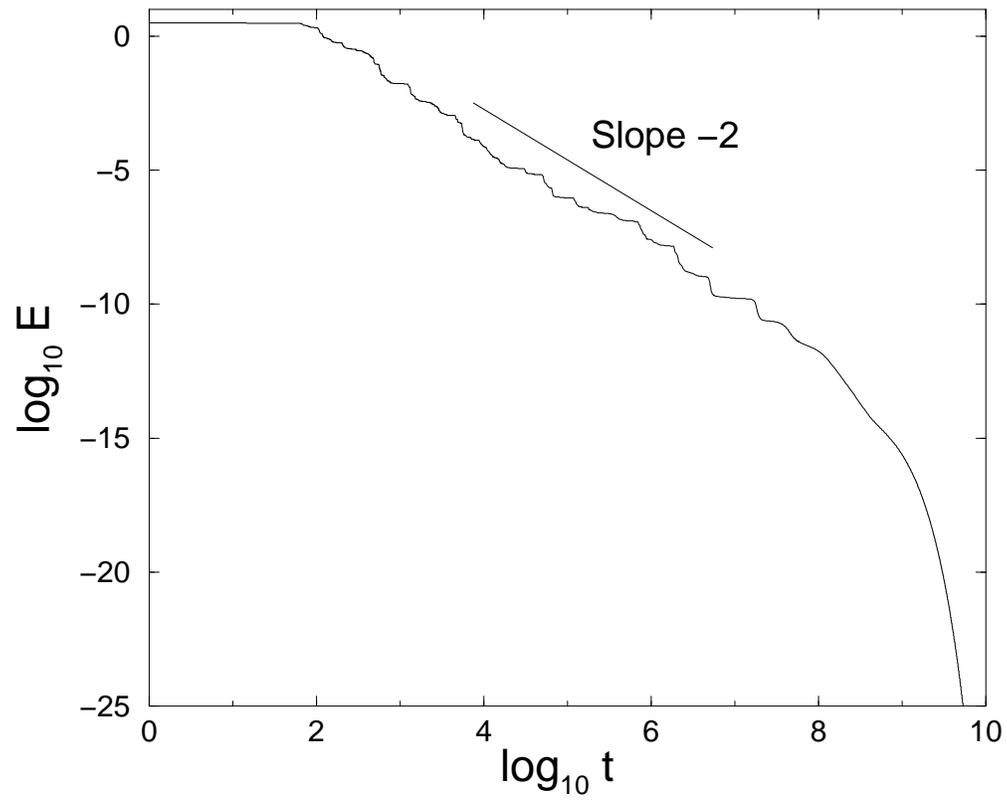}}}}
\caption{Time decay of the total energy $E(t)$ as a function of time $t$ 
for one particular realization. Note the step-wise, self-similar behavior. The average
slope $-2$ is also shown.}
\label{fig1}
\end{figure}

\begin{figure}
\epsfxsize=.8\hsize
{\hskip 0.0cm{\centerline{\epsfbox{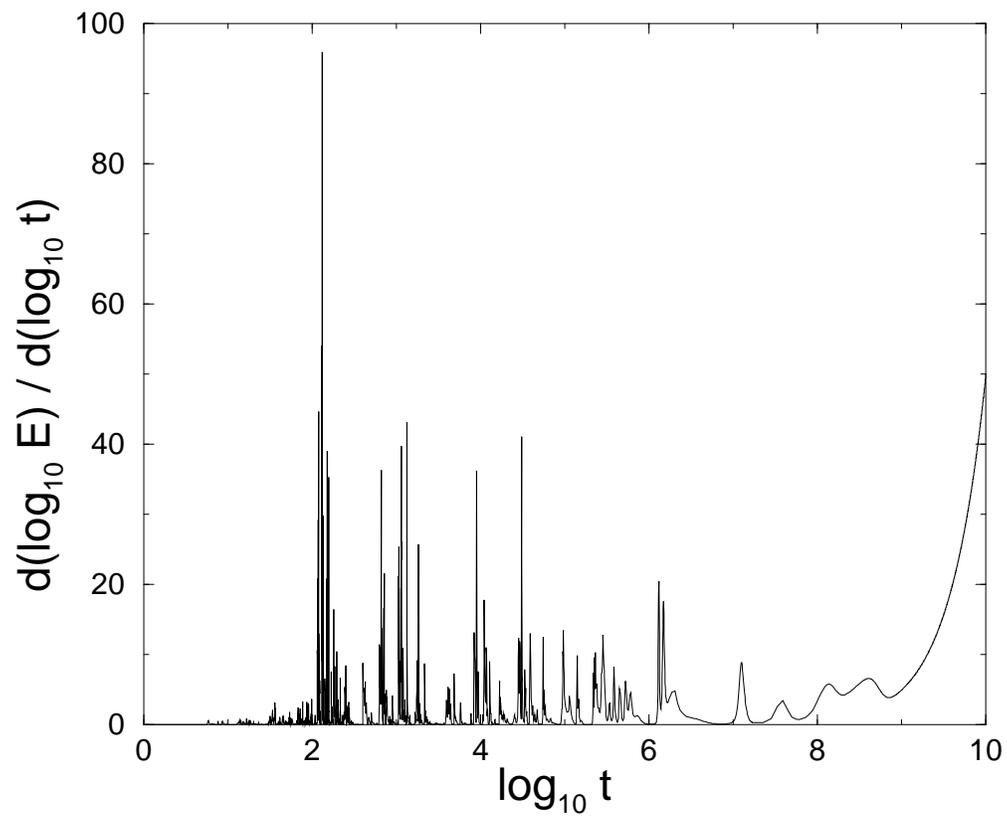}}}}
\caption{Time derivative of the energy in figure \ref{fig1}.}
\label{fig2}
\end{figure}

During the decay stage we measure the total energy 
$E(t) = \sum_{n=0}^{N-1} \left| u_n(t)\right|^2$
as a function of the decay time $t$ 
(the time elapsed from when we switched off the forcing). 
In figure \ref{fig1}  the decay of the total energy on a {\it single}
 run is shown, 
i.e., no ensemble average is performed here. 
The total energy decays in  burst, i.e., it is constant for some
time and than very suddenly drops.  
As this behavior occurs on all scales,
 the step-like structure looks self similar.
The averaged (over the whole decay) decay exponent is close to
$-2$. For very large times it decays exponentially.
The derivative of the energy  decay is shown in figure \ref{fig2} in 
order to better highlight the bursty structure. Also from figure \ref{fig2} 
one can immediately realize that the typical time scales 
during the decay becomes larger and larger.
The bursty structure of the energy decay in the GOY model has been
analyzed in detail by Okkels and Jensen \cite{okk97,okk00} and it reflects
the intermittent behavior of the energy flux downscale.

\begin{figure}
\epsfxsize=.8\hsize
{\hskip 0.0cm{\centerline{\epsfbox{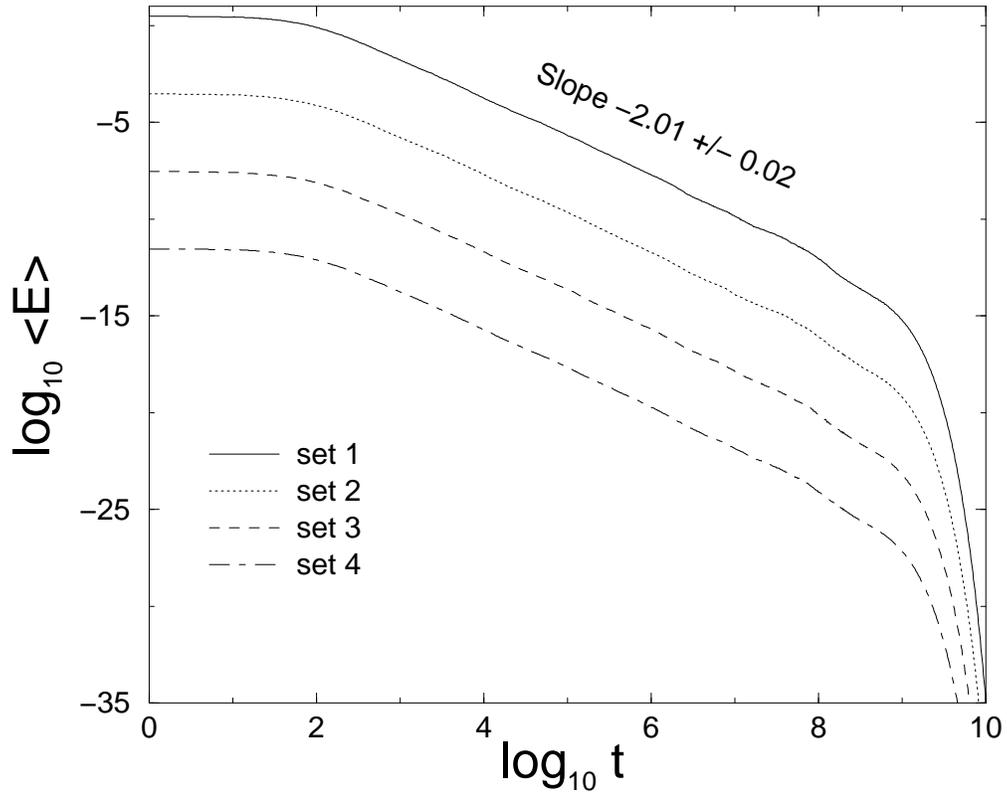}}}}
\caption{Ensemble averaged energy as a function of decay time. 
In order to give an idea about the statistical 
error, four  sets are shown (shifted in y-direction). Each one is obtained
by averaging 500 uncorrelated decay processes. }
\label{fig3}
\end{figure}

We now {\it ensemble average} $E(t)$ by collecting various
 starting configurations
as described above and letting them decay.
In figure \ref{fig3} we show the time decay of the {ensemble-averaged}
total energy. The step like behavior of a single realization of $E(t)$
is now completely smeared out and the decay law $\left< E(t) \right>
\sim t^{-2}$ is revealed. 
For large times the averaged decay is of course again
exponential.

We now
set up a simple model which is able to describe 
the {\it average} energy behavior during the decay process.
The model closely follows the mean-field model of reference \cite{loh94a}.
The major assumption we make is to suppose that all the energy is 
contained in the smallest shells corresponding to the largest scales. 
For simplicity we assume that it is only in the zeroth shell of eq.\
(\ref{goyeq}), 
$E\simeq \left< \left|u_0\right|^2\right>$.
During the first part of the decay energy will disappear from the
zeroth shell  by being  transferred 
to smaller scales $ {d\over dt} 
\left< { {\left| u_0\right|}^2}\right> 
\sim - \left<\left| u_0 \right|^3\right>$ and hence
${d\over dt}  \left< E(t)\right>  \sim - \left<E(t)\right>^{3/2}$
with the solution  
\begin{equation}
\left<E(t)\right>
=\left[E(t_0)^{-1/2}+{1\over 2}\left(t-t_0\right)\right]^{-2}.
\label{decaymode}
\end{equation}
Notice that asymptotically this means that, for $t\gg t_0$, 
we expect the energy to decay quadratically in time, $\left<E(t)\right>
\sim t^{-2}$, just as seen in figure \ref{fig3}. 
The reason that the mean field argument works for the 
averaged decay is that the required time for the transport
of pulses downscale is determined by the large scale dynamics,
see e.g.\ figure 1b of Sch\"orghofer \cite{sch98}.

As energy is removed from the system the effective Reynolds number will  
decrease further and further. In particular there will come a time, $t_1$, 
for which even on the zeroth shell  the dissipative term will dominate with respect to the non linear term in equation (\ref{goyeq}).
From that moment on, the equation for the energy decay will be  
${1\over 2} \left< {\dot {\left| u_0\right|}^2}\right> \simeq - \nu k_0^2 
\left< \left| u_0 \right|^2\right>$ and hence
$
{d\over dt} \left< E(t)\right> \sim - \nu k_0^2 \left<E(t)\right>
$
whose solution is an exponential damping,
\begin{equation}
\left< E(t)\right> 
=E(t_1)\cdot \exp{\left[-2\nu k_0^2\left(t-t_1\right)\right]}.
\label{exp_decay}
\end{equation}
As seen from figure \ref{fig3}, equations (\ref{decaymode}) and
(\ref{exp_decay}) correctly describe the respective
short and long term behavior
of the energy decay.

\section{Intermittency in decaying turbulence}
The time behavior of an individual decay process 
is  complicated because of the bursty 
structure of the decay, see figure \ref{fig2}. The presence of bursts is
 an essential feature of intermittency. 
During the decay process both the intensities of the bursts decrease and 
their duration increase  orders of magnitudes.

The question now is 
if or not the statistical properties of the turbulent fluctuations
remain the same during the decay process.
To answer this question we study the k-scaling of 
higher order moments. 
Rather than calculating the scaling of moments of the velocity $u_n$ itself, 
which 
within the GOY model shows unphysical period 2 and period 3 oscillations 
\cite{kad95}, we focus on the scaling of
energy flux moments \cite{kad95}  $\Sigma_{n,p}$,
\begin{equation}
\Sigma_{n,p}(t)=\left\langle \left| {\cal I}
\left( u_n u_{n+1} u_{n+2}+{1\over 4}u_{n-1} u_{n} u_{n+1}\right) 
\right|^p \right\rangle .
\end{equation}
Here $\cal I$ denotes the imaginary part.
The $\Sigma_{n,p}$ are free of the period 2 and
period 3 artifacts in the spectrum and show very clean scaling properties 
\cite{kad95}.
With the angular brackets $\left\langle\dots\right\rangle$ we again
denote ensemble average conditioned to a given decay time $t$. 
However, to obtain better statistics, here we had in addition to 
average over a short period of time. We chose a tenth of  a decade
and the given times refer to the {\it end} of that small time interval.

\begin{figure}
\epsfxsize=.8\hsize
{\hskip 0.0cm{\centerline{\epsfbox{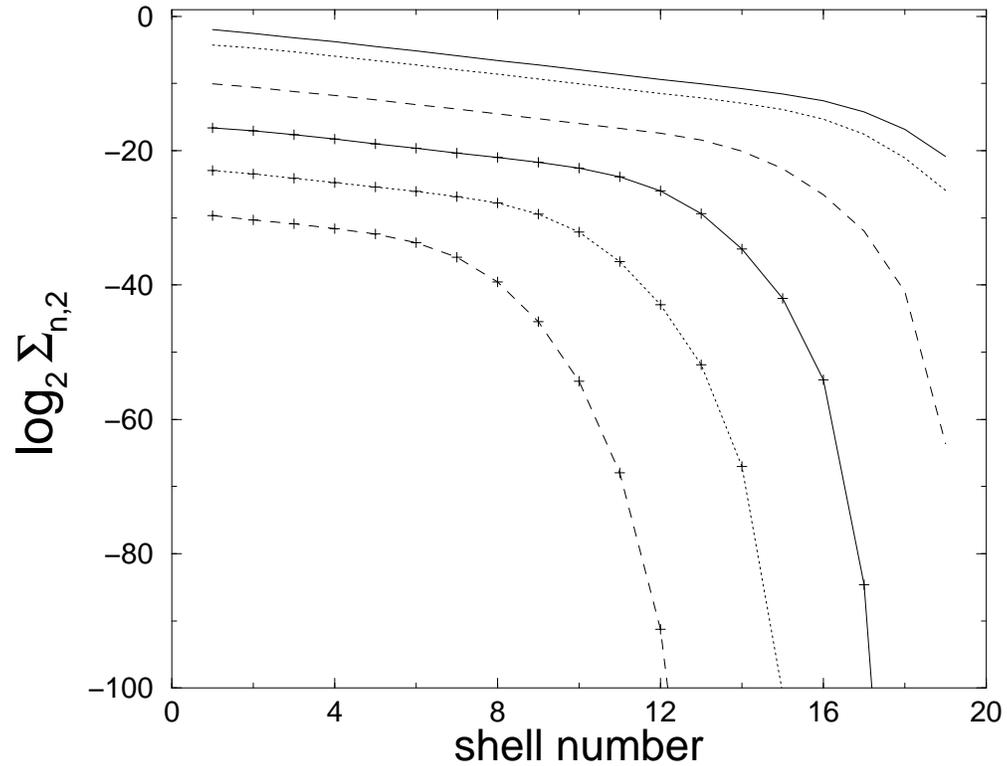}}}}
\caption{Second order flux moment $\Sigma_{n,2}$ at six
different decay times. The upper curve is the original situation,
the following curves reflect the situation one, two, three, four, and 
five decades in time later. Besides ensemble averaging over 500 ensembles,
the respective data are also averaged over one tenth of a decade in time.}
\label{fig4}
\end{figure}

The numerical results are shown in figure \ref{fig4} where we have plotted 
various moments of the fluxes as a function of decay time.
One evident feature is the decrease of the Reynolds number
 as the decay goes on. This 
results
from the strong increase of the dissipative scale and consequently 
the shortening of the inertial range.
In the remaining inertial subrange (ISR) the slope looks very similar,
but whether it really is the same cannot yet be judged from this type
of plot.

\begin{figure}
\epsfxsize=.8\hsize
{\hskip 0.0cm{\centerline{\epsfbox{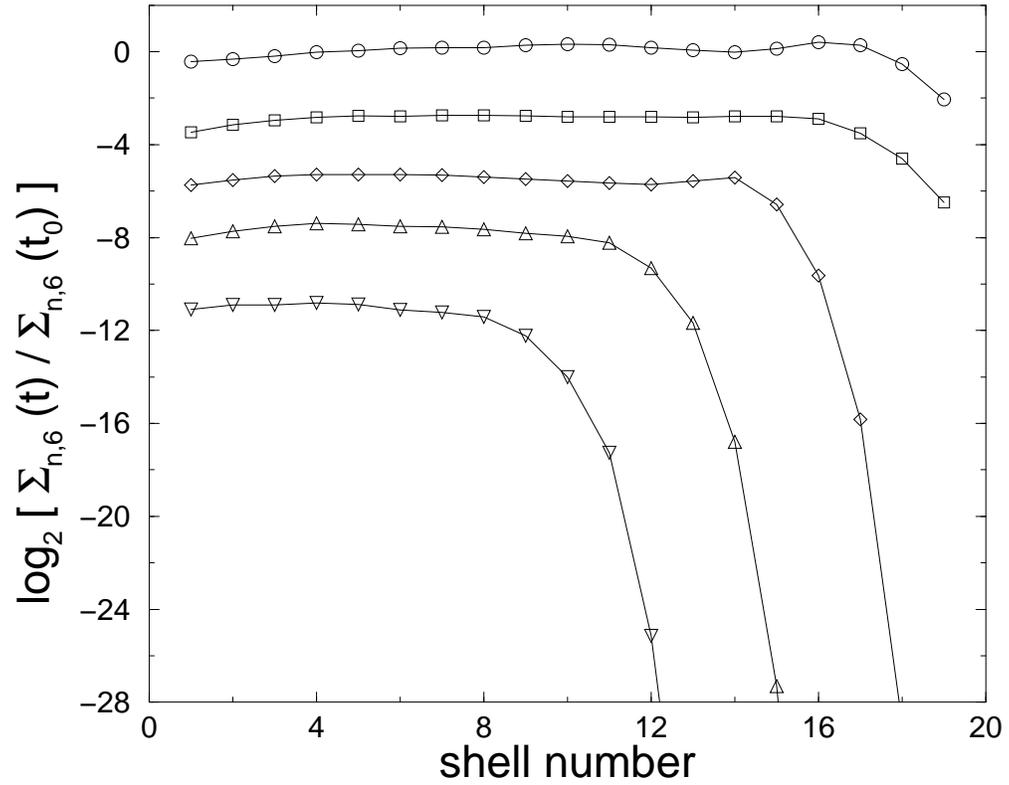}}}}
\caption{Ratio  of $\Sigma_{n,6}(t) $ to $\Sigma_{n,6}(t_0) $  
one, two, three, four, and five decades after $t_0$ when 
the decay started. For clarity the curves are again
shifted in y-direction.}
\label{fig5}
\end{figure}

Therefore, in order to explore whether the scaling in the forced and in the
unforced cases are really the same,
we calculate the ratio between powers of the 
fluxes computed at different decay times, see figure \ref{fig5}.
Though in the forced case the forcing is limited to the zeroth shell,
slight deviations spread over the first three shells or so. 
However, in the ISR
the scaling properties and thus the
intermittency really seem to be  the same, 
within statistical errors. 
Note that getting this conclusion from DNS would be very hard, due
to the small extension of the inertial subrange. 

This result on the lack of dependence of intermittency on the forcing
resembles analogous conclusions from analysing the effect
of the viscous subrange on the ISR scaling: For both the GOY model and
also for DNS She and coworkers \cite{lev95,cao96} found the same degree of
ISR intermittency, independently of whether normal or hyperviscosity was
employed. Only slightly beyond the onset of turbulence there may be
a small dependence on the type of viscosity within the GOY model \cite{sch95}.

\section{Kicked turbulence}
Kicked turbulence has been analysed within the framework of a
 mean-field theory \cite{loh00}.
In the kicked case the GOY model is forced with a delta-like (in time)  
 forcing $g \delta_{n,0}$  with  frequency $f$.
With delta-like forcing we mean that the forcing $g \delta_{n,0}$ 
is periodically turned on for a small time $\Delta t_{kick} \ll f^{-1}$.
The presence of the forcing will sustain the energy flux. 
The  turbulence level achieved in kicked turbulence depends on 
both the forcing strength 
$A= g\Delta t_{kick}$ 
and the forcing frequency $f$. We employ the
GOY model dynamics to explore  
this dependency beyond the simple mean field approach of reference
\cite{loh00}.

\begin{figure}
\epsfxsize=1.12\hsize
{\hskip 0.0cm{\centerline{\epsfbox{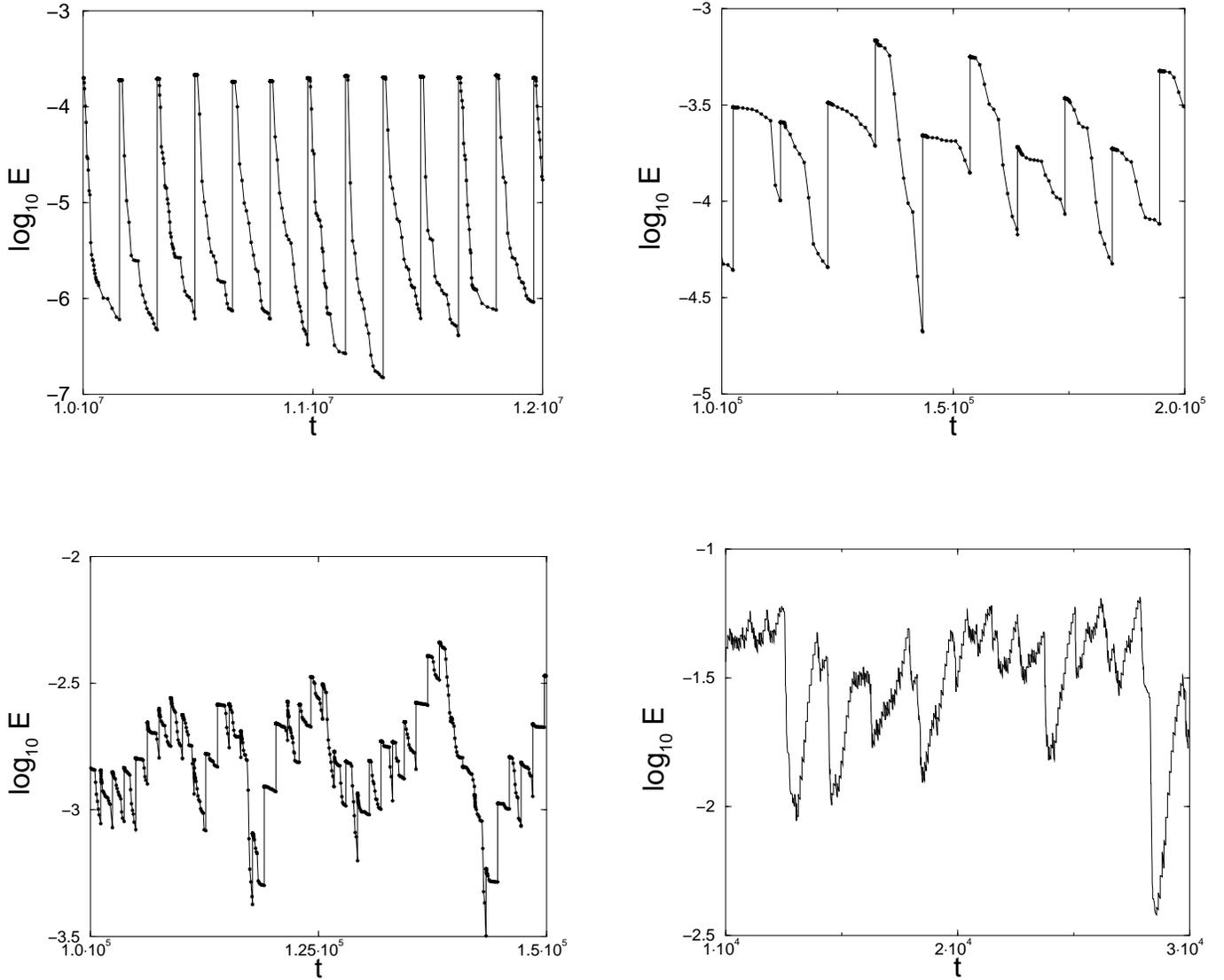}}}}
\caption{(Top-Left): Total energy in the GOY as a function of time, 
in the laminar regime 
($f=6\cdot 10^{-6}$, $A=0.01$).
(Top-Right): Closely after the laminar regime
($f=1\cdot 10^{-4}$, $A=0.01$).
(Bottom-Left): Close to the turbulent regime. 
($f=1\cdot 10^{-3}$, $A=0.01$).
(Bottom-Right): In the turbulent regime.
($f=1.5 \cdot 10^{-1}$, $A=0.01$).}
\label{fig8}
\end{figure}

The qualitative behavior of the energy, as a  function 
of kicking frequency can be seen from figures \ref{fig8}.
 In figure \ref{fig8}a we can see the energy behavior in 
the laminar regime. After each kick the system jumps at 
(almost) constant upper levels. From this value it then 
decays for a long period before another kick is applied.

In figure \ref{fig8}b we are in the transition regime towards turbulence. 
The time between two kicks has been decreased as compared to
 Fig.\ \ref{fig8}a and hence the system does not have sufficient 
time to fully relax. 
The kicks still heavily influence the macroscopic behavior.

In figure \ref{fig8}c the kicking frequency has been further increased and
we are close to the turbulent regime. 
We see regions where the energy starts to pile up for a while, 
after which it relaxes very fast through bursts of energy. 
The macroscopic energy behavior is described by a competition 
between the kicking (with timescale $1/f$)  
and the energy decay between the kicks (with the time scale of the
large eddy turnover time).

In figure \ref{fig8}d we are in the fully turbulent regime. 
Here the individual kicks are no longer important  for the macroscopic 
energy behavior and they acts as a sort of average constant forcing.
Energy can build up over many kicking timescales before it is released
through an energy burst, whenever the phase relation is appropriate 
\cite{okk97}. This behavior is the analog to the bursty structure
of figures \ref{fig1} and \ref{fig2} in the decaying case: Energy can only
be transported downscale if the phase relations happen to be appropriate
\cite{okk97}. If not, an energy plateau forms in the decaying case
or the energy piles up in the kicked case.

\begin{figure}
\epsfxsize=.8\hsize
{\hskip 0.0cm{\centerline{\epsfbox{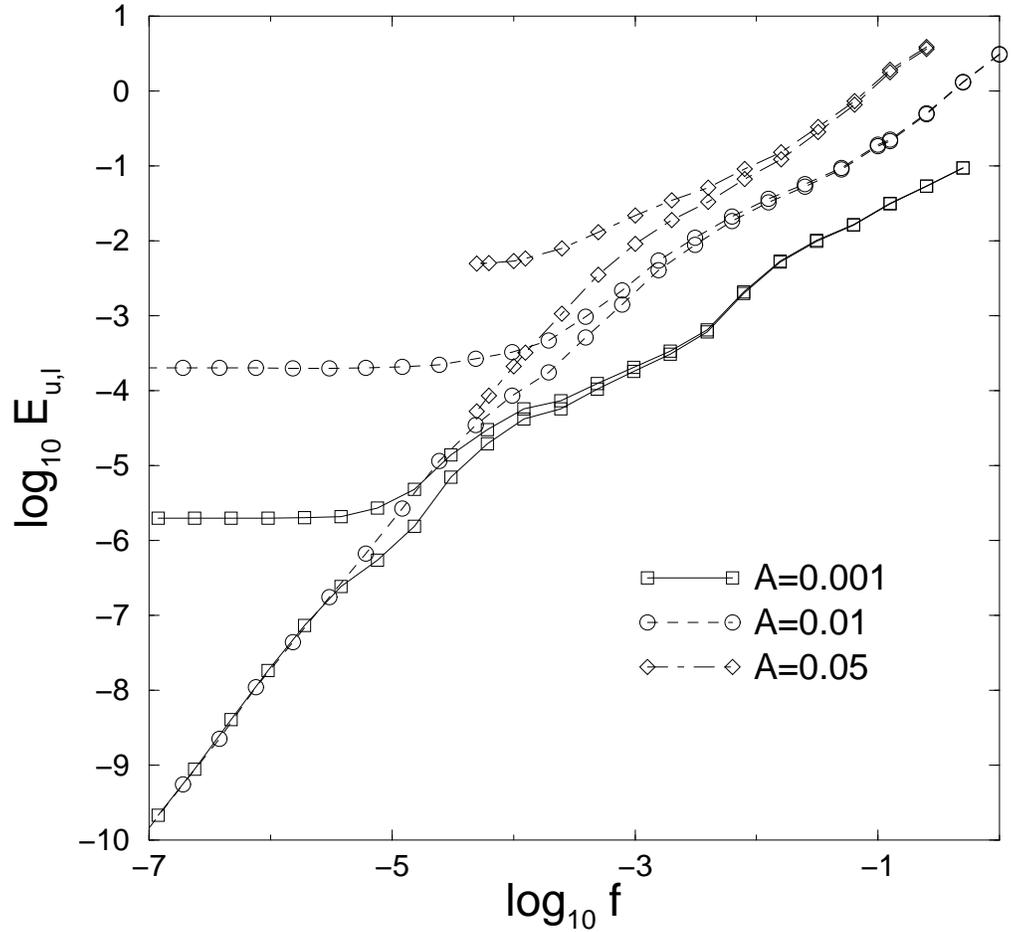}}}}
\caption{Numerical computation of the energy levels. $E_l$ (lower curves) 
and $E_u$ (upper curves) as a function of the frequency $f$ for three 
different kicking strengths $A$. For large $f$ the slope is roughly 
$E_u\sim E_l \sim f$.}
\label{fig7}
\end{figure}

To get the dependence of the total energy just before the kick ($E_l$)
and just after the kick ($E_u$) as a function of the kicking frequency
and the kicking strength we again performed ensemble averages of many
realizations.  
In figure \ref{fig7} we show 
 $E_l$ and $E_u$  as a function of $f$ for three different 
forcing strengths $A$.
Just as in the mean field 
case \cite{loh00} three regimes can be seen: 
(i) A laminar regime in which the upper energy level is constant
and corresponds to the total energy input during the kick and where
the time between the kicks is sufficiently large so that the 
energy completely decays;
(ii) a transitional regime; and 
(iii) a turbulent regime where the average 
upper and lower energy level are  equal
 as the forcing is experienced
as a continuous forcing. The energy levels roughly scale with 
$f^1$. 
In the last regime the features are different from those
of the third regime in the mean field theory \cite{loh00} where the
 average energy of the lower level is always less than that of the
upper level. The reason is that in the dynamical model energy can
build up over many forcing periods due to phase blocking. Therefore,
it is experienced as a continuous forcing.

Nevertheless, at least 
the scaling can be obtained by a similar argument
as employed in the mean field theory \cite{loh00}:
If the energy of the GOY can be (roughly) approximated by the 
energy contained in the largest shell, $E_l\sim \left<|u_0|^2\right>$, then 
directly after the kick  it will be $E_u 
\sim \left<|u_0+ g\Delta t_{kick} |^2\right>
= \left<|u_0+A|^2\right>
=
E_l + 2 A \sqrt{E_l} + A^2.
$
Between two kicks the system is freely decaying. Therefore
we can apply in between two kicks the mean field result of refs.\
\cite{loh94a,loh00} to  
connect $E_u$ and $E_l$.
If the decay starts from an energy value $E_u$, after a time $1/f$ ,
when the next kick is applied,
the energy will be decayed to $E_l$. The two energy levels
are connected through the equation \cite{loh94a,loh00}
\begin{equation}
{1 \over {f\tau}} = 3 \left[ F(Re(E_l)) - F(Re(E_u))\right] .
\label{eqnlohse}
\end{equation}
Here the function $F(Re)$ is defined as \cite{loh94a}
\begin{equation}
F(Re)={1 \over {2 Re^2}} \left\{ -\gamma + \sqrt{\gamma^2 +Re^2}\right\} + {1 \over {2\gamma}}\left\{ {{\gamma+\sqrt{\gamma^2+Re^2}} \over {Re}}\right\}
\label{def_f}
\end{equation}
with $\gamma =9$. The
Reynolds number and the energy in the GOY are connected 
by 
$Re(E(t))= \sqrt{2 \over 3}{{L\sqrt{E(t)}} \over {\nu}}$.
Here, 
$L$ is the integral 
scale and $\tau=L^2/{\nu}$ a viscous time scale.

\begin{figure}
\epsfxsize=.8\hsize
{\hskip 0.0cm{\centerline{\epsfbox{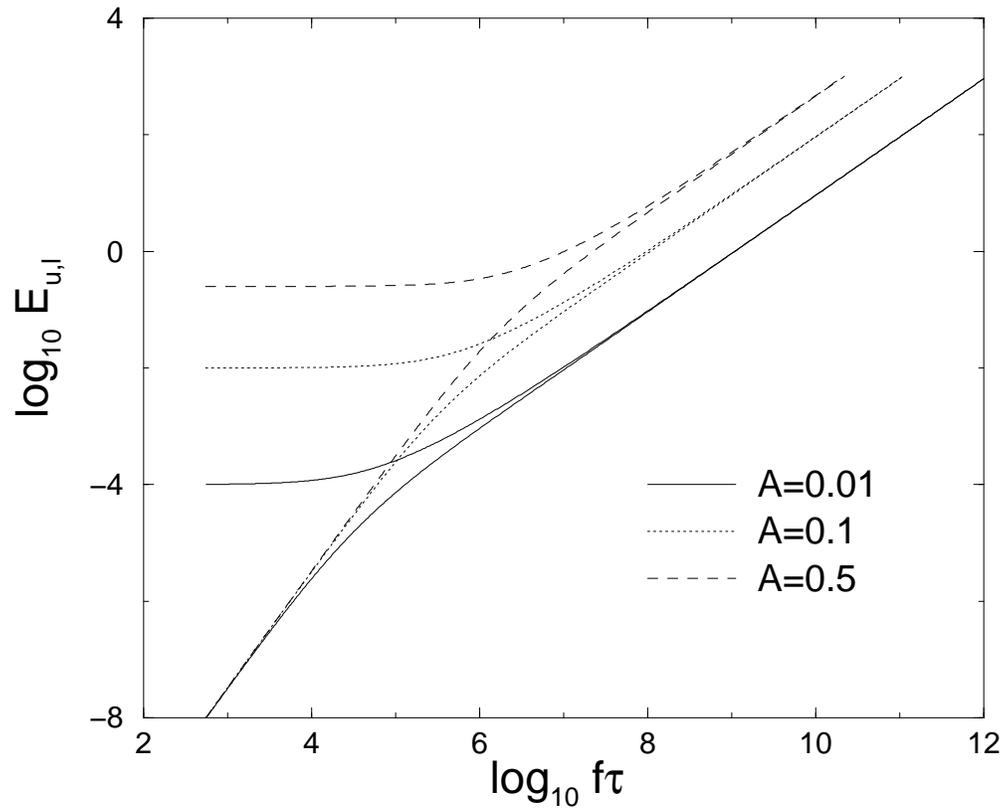}}}}
\caption{Energy levels $E_u$ (upper curves) and $E_l$ (lower curves) 
as a function of the frequency for three different kicking strength,
as they follow from the dimensional argument given in the text.}
\label{fig6}
\end{figure}

Solving equation (\ref{eqnlohse}) we can find the value of the 
upper, $E_u$, and lower, $E_l$, energy levels as a function 
of the forcing strength $A$ and frequency $f$.
In figure \ref{fig6} this dependence is plotted, revealing the 
basic features as in the numerical figure \ref{fig7}.
For large $Re \gg \gamma$, we have $F(Re) =1/Re$ and one finds
$E_u\sim E_l \sim f$, just as in the numerical case.

\section{Conclusions}
We summarize our main findings:

The GOY model is employed to study both decaying and periodically
kicked turbulence. 
Energy is transfered towards the small scales in intermittent
{\it bursts}, leading to stepwise behavior in the decaying
case or energy pileups and subsequent bursts in the kicked case.
In spite of this intermittent behavior, 
mean field arguments as developed in refs.\cite{loh94a,loh00} 
are sufficient to 
account for the ensemble averaged energy decay $\left<E(t)\right>
 \sim t^{-2}$ 
or the parameter dependence $E_u \sim E_l \sim f$ 
for the ensemble averaged total energy
in the kicked case. 
The reason that mean field arguments work here 
is that the decay and the kicking is 
determined by the large scale dynamics.

For what concerns the statistical properties of decaying turbulence, 
our finding supports the idea that decaying turbulence has the same 
intermittency as stationary turbulence. 
In particular this finding allows to conclude that at least in the
GOY model  ISR intermittency is independent on the  
 forcing mechanism of turbulence.
The only relevant dynamics during the decay process seems to be a 
shortening  of the inertial range (decrease of Reynolds number) 
but leaving  the same intermittency properties.

\acknowledgments
The authors thank L. Biferale for fruithfull discussions.
This work is  part of the research  programme of
the ``Stichting voor Fundamenteel Onderzoek der Materie (FOM)'', which
is    financially supported   by  the  ``Nederlandse  Organisatie voor
Wetenschappelijk Onderzoek (NWO).
This research was also supported in part 
by the European Union under contract HPRN-CT-2000-00162 and
by the German-Israeli Foundation (GIF).

%\bibliography{literatur}

\vspace{1cm}
\noindent
$^*$ Corresponding author

%%%%%%%%%%%%%%%%%%%% Figure 1 %%%%%%%%%%%%%%%%%%%%%%%%%%%%%%%%%%%%%%%%%%%%%%

%%%%%%%%%%%%%%%%%%%% Figure 2 %%%%%%%%%%%%%%%%%%%%%%%%%%%%%%%%%%%%%%%%%%%%%%

%%%%%%%%%%%%%%%%%%%% Figure 3 %%%%%%%%%%%%%%%%%%%%%%%%%%%%%%%%%%%%%%%%%%%%%%

%%%%%%%%%%%%%%%%%%%% Figure 4 %%%%%%%%%%%%%%%%%%%%%%%%%%%%%%%%%%%%%%%%%%%%%%

%%%%%%%%%%%%%%%%%%%% Figure 5 %%%%%%%%%%%%%%%%%%%%%%%%%%%%%%%%%%%%%%%%%%%%%%

%%%%%%%%%%%%%%%%%%%% Figure 7 %%%%%%%%%%%%%%%%%%%%%%%%%%%%%%%%%%%%%%%%%%%%%%

%%%%%%%%%%%%%%%%%%%% Figure 6 %%%%%%%%%%%%%%%%%%%%%%%%%%%%%%%%%%%%%%%%%%%%%%

\end{document}